\documentclass[pra,twocolumn,preprintnumbers,showpacs,
nofootinbib,floatfix]{revtex4-1}

\usepackage{graphicx,bm,amsmath}
\usepackage{bm}
\usepackage{tikz}
\usepackage[sort&compress]{natbib}

\makeatletter

\def\graphicscale{\twocolumn@sw{0.30}{0.33}}
\def\graphicthreescale{\twocolumn@sw{0.30}{0.33}}

\makeatother

\begin{document}

\def\blue{\color{blue}}
\def\cyan{\color{cyan}}
\def\black{\color{black}}
\def\green{\color{green}}
\def\magenta{\color{magenta}}
\def\mahogany{\color{brown}}
\def\red{\color{red}}
\def\orange{\color{orange}}
\def\purple{\color{purple}}
\def\violet{\color{violet}}
\def\teal{\color{teal}}
\def\pink{\color{pink}}
\def\lime{\color{lime}}
\def\yellow{\color{yellow}}
\def\olive{\color{olive}}

\newcommand{\airplane}{\ding{40}}
\newcommand{\fullhand}{\ding{42}}
\newcommand{\writehand}{\ding{45}}
\newcommand{\pencildown}{\ding{46}}
\newcommand{\pencil}{\ding{47}}
\newcommand{\pencilup}{\ding{48}}
\newcommand{\yes}{\ding{51}}
\newcommand{\bigyes}{\ding{52}}
\newcommand{\no}{\ding{55}}
\newcommand{\bigno}{\ding{56}}
\newcommand{\arrow}{\ding{243}}
\newcommand{\hand}{\ding{43}}
\newcommand{\largearrow}{\ding{224}}
\newcommand{\altarrow}{\ding{219}}

\def\reff#1{{\footnotesize\mahogany #1}}

\newcommand{\lsim}{ {\
\lower-1.2pt\vbox{\hbox{\rlap{$<$}\lower5pt\vbox{\hbox{$\sim$}}}}\ } }
\newcommand{\gsim}{ {\
\lower-1.2pt\vbox{\hbox{\rlap{$>$}\lower5pt\vbox{\hbox{$\sim$}}}}\ } }

\def\dof{\mathrm{dof}}
\def\er#1#2{\relax\ifmmode{}^{+#1}_{-#2}\else$^{+#1}_{-#2}$\fi}
\def\erparen#1#2{\relax\ifmmode{}(^{#1}_{#2})\else$(^{#1}_{#2})$\fi}
%

\def~{\ifmmode\phantom{0}\else\penalty10000\ \fi}
\def\half{{1\over2}}
\def\bra#1{\left\langle #1\right|}
\def\ket#1{\left| #1\right\rangle}
\def\vev#1{\left\langle #1\right\rangle}
\def\hc{\mathrm{h.c.}}
\def\su#1{{SU\left(#1\right)}}
\def\u#1{{U\left(#1\right)}}
\def\d#1#2{d\mskip 1.5mu^{#1}\mkern-1mu{#2}\,}
\def\D#1#2{{\d#1{#2} \over (2\pi)^{#1}}\,}
\def\dtilde#1{{\d3{#1} \over (2\pi)^3 \,2\omega_{#1}}\,}
\def\tr{\,\mathrm{tr}}
\def\Tr{\,\mathrm{Tr}}
\def\str{\,\mathrm{str}}
\def\det{\,\mathrm{det}}
\def\sdet{\,\mathrm{sdet}}
\def\fm{\mathrm{fm}}
\def\ev{\mathrm{e\kern-0.1em V}}
\def\kev{\mathrm{ke\kern-0.1em V}}
\def\mev{\mathrm{Me\kern-0.1em V}}
\def\gev{\mathrm{Ge\kern-0.1em V}}
\def\tev{\mathrm{Te\kern-0.1em V}}
\def\ps{\mathrm{ps}}
\def\re{\,\mathrm{Re}}
\def\im{\,\mathrm{Im}}
\let\Re=\re \let\Im=\im
\def\dotp#1#2{#1\mathord\cdot #2}
\def\n#1e#2n{{#1}\times 10^{#2}}
\def\mb{\mathrm{MB}}
\def\gb{\mathrm{GB}}
\def\tb{\mathrm{TB}}

\def\ra{\rangle}
\def\la{\langle}
\def\ord#1{\mathcal{O}(#1)}
\def\bea{\begin{eqnarray}}
\def\eea{\end{eqnarray}}
\def\nn{\nonumber}
\def\On{\mathcal{O}_n}

\def\cO{\mathcal{O}}
\def\cF{\mathcal{F}}
\def\cP{\mathcal{P}}
\def\cH{\mathcal{H}}
\def\cL{\mathcal{L}}
\def\cM{\mathcal{M}}
\def\cD{\mathcal{D}}
\def\cE{\mathcal{E}}
\def\cR{\mathcal{R}}
\def\cpv{
\math{\mathrm{CP}\hspace{-0.6cm}\slash{}\hspace{0.4cm}}}
\def\ods2{\mathcal{O}_{\Delta S=2}}
\def\zds2{Z_{\Delta S=2}}
\def\orda{$\blue O(a)$}

\def\sec{\mathrm{s}}
\def\K{\mathrm{K}}
\def\msbar{{\overline{\mathrm{MS}}}}
\def\NDR{\mathrm{NDR}}
\def\RI{\mathrm{RI}}
\def\MOM{\mathrm{MOM}}
\def\RGI{\mathrm{RGI}}
\def\BSM{\mathrm{BSM}}
\def\LO{\mathrm{LO}}
\def\NLO{\mathrm{NLO}}
\def\NNLO{\mathrm{NNLO}}
\def\AWI{\mathrm{AWI}}
\def\VWI{\mathrm{VWI}}
\def\W{\mathrm{W}}
\def\PCAC{\mathrm{PCAC}}
\def\bare{\mathrm{bare}}
\def\crit{\mathrm{crit}}
\def\sub{\mathrm{sub}}
\def\lqcd{\Lambda_\mathrm{QCD}}
\def\lchi{\Lambda_\chi}
\def\lchitwo{\Lambda_\chi^{N_f=2}}
\def\lchithree{\Lambda_\chi^{N_f=3}}
\def\lat{\mathrm{lat}}
\def\phys{\mathrm{phys}}
\def\expt{\mathrm{expt}}
\def\max{\mathrm{max}}
\def\min{\mathrm{min}}
\def\sea{\mathrm{sea}}
\def\val{\mathrm{val}}
\def\qcd{\mathrm{QCD}}
\def\SI{\mathrm{SI}}
\def\SD{\mathrm{SD}}

\def\nuc#1#2#3{\phantom{|}^{#2}_{#3}#1}

\def\Metafoot{M_{\mbox{\footnotesize ``$\eta$''}}}
\def\Metascript{M_{\mbox{\scriptsize ``$\eta$''}}}
\def\Metatiny{M_{\mbox{\tiny``$\eta$''}}}

\def\spqr{SPQ$_{cd}$R}

\def\qqbarind#1#2{^#1_{\phantom{#1}#2}}

\makeatletter
\def\slash#1{{\mathpalette\c@ncel{#1}}} 
\def\big#1{{\hbox{$\left#1\vbox to1.012\ht\strutbox{}\right.\n@space$}}}
\def\Big#1{{\hbox{$\left#1\vbox to1.369\ht\strutbox{}\right.\n@space$}}}
\def\bigg#1{{\hbox{$\left#1\vbox to1.726\ht\strutbox{}\right.\n@space$}}}
\def\Bigg#1{{\hbox{$\left#1\vbox
to2.083\ht\strutbox{}\right.\n@space$}}}
\makeatother

\newcommand{\rC}{C}
\newcommand{\gA}{A}
\newcommand{\oP}{P} 

\def\good{\raisebox{0.35mm}{{\color{green}$\bigstar$}}}
\def\bad{\raisebox{0.35mm}{\hspace{0.65mm}{\color{red}\tiny$\blacksquare$}}} 
\def\soso{\hspace{0.25mm}\raisebox{-0.2mm}{{\color{green}\Large$\circ$}}}
\def\okay{\hspace{0.25mm}\raisebox{-0.2mm}{{\color{green}\large\checkmark}}}

\newcommand\leftidx[3]{%
  {\vphantom{#2}}#1#2#3%
}

\def\Hone{\leftidx{^1}{\mathrm{H}}{}}
\def\Htwo{\leftidx{^2}{\mathrm{H}}{}}
\def\Hthree{\leftidx{^3}{\mathrm{H}}{}}
\def\Hethree{\leftidx{^3}{\mathrm{He}}{}}
\def\Hefour{\leftidx{^4}{\mathrm{H}}{}}

\def\Mss{M_{\bar{s}s}}
\def\cut{\mathrm{cut}}

\def\sigrel{\sigma^\mathrm{rel}}

\title{Lattice computation of the nucleon scalar quark contents at the physical point}

 \author{S.~Durr$^{1,2}$, Z.~Fodor$^{1,2}$, C.~Hoelbling$^1$, S.~D.~Katz$^{3,4}$, S.~Krieg$^{1,2}$, L.~Lellouch$^5$, T.~Lippert$^{1,2}$, T.~Metivet$^{5,6}$, A.~Portelli$^{5,7}$, K.~K.~Szabo$^{1,2}$, C.~Torrero$^5$, B.~C.~Toth$^1$, L.~Varnhorst$^1$}
 \affiliation{$^1$Department of Physics, University of Wuppertal, D-42119 Wuppertal, Germany}
\affiliation{ $^2$J\"ulich Supercomputing Centre, Forschungszentrum J\"ulich, D-52428 J\"ulich, Germany}
\affiliation{$^3$Institute for Theoretical Physics, E\"otv\"os University, H-1117 Budapest, Hungary}
\affiliation{$^4$MTA-ELTE Lend\"ulet Lattice Gauge Theory Research Group, H-1117 Budapest, Hungary}
\affiliation{$^5$CNRS, Aix-Marseille U., U. de Toulon, Centre de Physique Th\'eorique, UMR 7332, F-13288 Marseille, France}
\affiliation{$^6$CEA-Saclay, IRFU/SPhN,
91191 Gif-sur-Yvette France}
\affiliation{$^7$Higgs Centre for Theoretical Physics, School of Physics and Astronomy, The University of Edinburgh, Edinburgh EH9 3FD, UK}
\collaboration{Budapest-Marseille-Wuppertal collaboration}
 \date{\today}

 \begin{abstract} 

We present a QCD calculation of the $u$, $d$ and $s$ scalar quark
contents of nucleons based on 47 lattice ensembles with $N_f = 2+1$
dynamical sea quarks, 5 lattice spacings down to $0.054\,\text{fm}$,
lattice sizes up to $6\,\text{fm}$ and pion masses down to
$120\,\text{MeV}$. Using the Feynman-Hellmann theorem, we obtain
$f^N_{ud} = 0.0405(40)(35)$ and $f^N_s = 0.113(45)(40)$, which
translates into $\sigma_{\pi N}=38(3)(3)\,\text{MeV}$,
$\sigma_{sN}=105(41)(37)\,\text{MeV}$ and $y_N=0.20(8)(8)$ for the
sigma terms and the related ratio, where the first errors are
statistical and the second are systematic. Using isospin relations, we
also compute the individual up and down quark contents of the proton
and neutron (results in the main text).
\end{abstract}


\maketitle


\emph{Introduction -} The scalar quark contents of nucleons, $N$, are important properties of these particles that are conveniently parametrized by the dimensionless ratios~\footnote{We
  use the relativistic normalization $\langle
  N(\vec{p}^{\,\prime})|N(\vec{p})\rangle=2E_{\vec{p}}\, (2\pi)^3
  \delta^{(3)}(\vec{p}^{\,\prime}-\vec{p})$. With unit normalization,
  the r.h.s. would be multiplied by a factor $2 M_N$.}  
\begin{equation}
\label{definition}
\begin{split}
f_{ud}^N &= m_{ud}\frac{\langle N|\bar{u}u+\bar{d}d|N\rangle}{2M_N^2}\equiv\frac{\sigma_{\pi N}}{M_N}\ ,\\  
f_{q}^N &= m_{q}\frac{\langle N|\bar{q}q|N\rangle}{2M_N^2}\equiv\frac{\sigma_{qN}}{M_N}\ ,
\end{split}
\end{equation}  
where $N$ can be either a proton, $p$, or a neutron, $n$, at rest, the quark
field $q=u,d$, or $s$ and $m_{ud}=(m_u+m_d)/2$ is the average $u$-$d$
quark mass. Note that in the isospin limit, $m_u=m_d$,
$f_{ud}^n=f_{ud}^p$ and $f_s^n=f_s^p$ and we will generically call
these quantities $f_{ud}^N$ and $f_s^N$, respectively. Although they
cannot directly be accessed in experiment, they are scheme and
scale-independent quantities that allow to translate quark-level
couplings into effective, scalar couplings with a nucleon. They are
related to a wide variety of observables such as pion and kaon-nucleon
scattering amplitudes, quark-mass ratios or quark-mass contributions
to nucleon masses. Their knowledge is also very important for Dark
Matter (DM) searches, as they allow to convert DM-quark couplings into
spin-independent, DM-nucleon cross sections.

\begin{figure*}[ht!]
\includegraphics[width=0.48\textwidth]{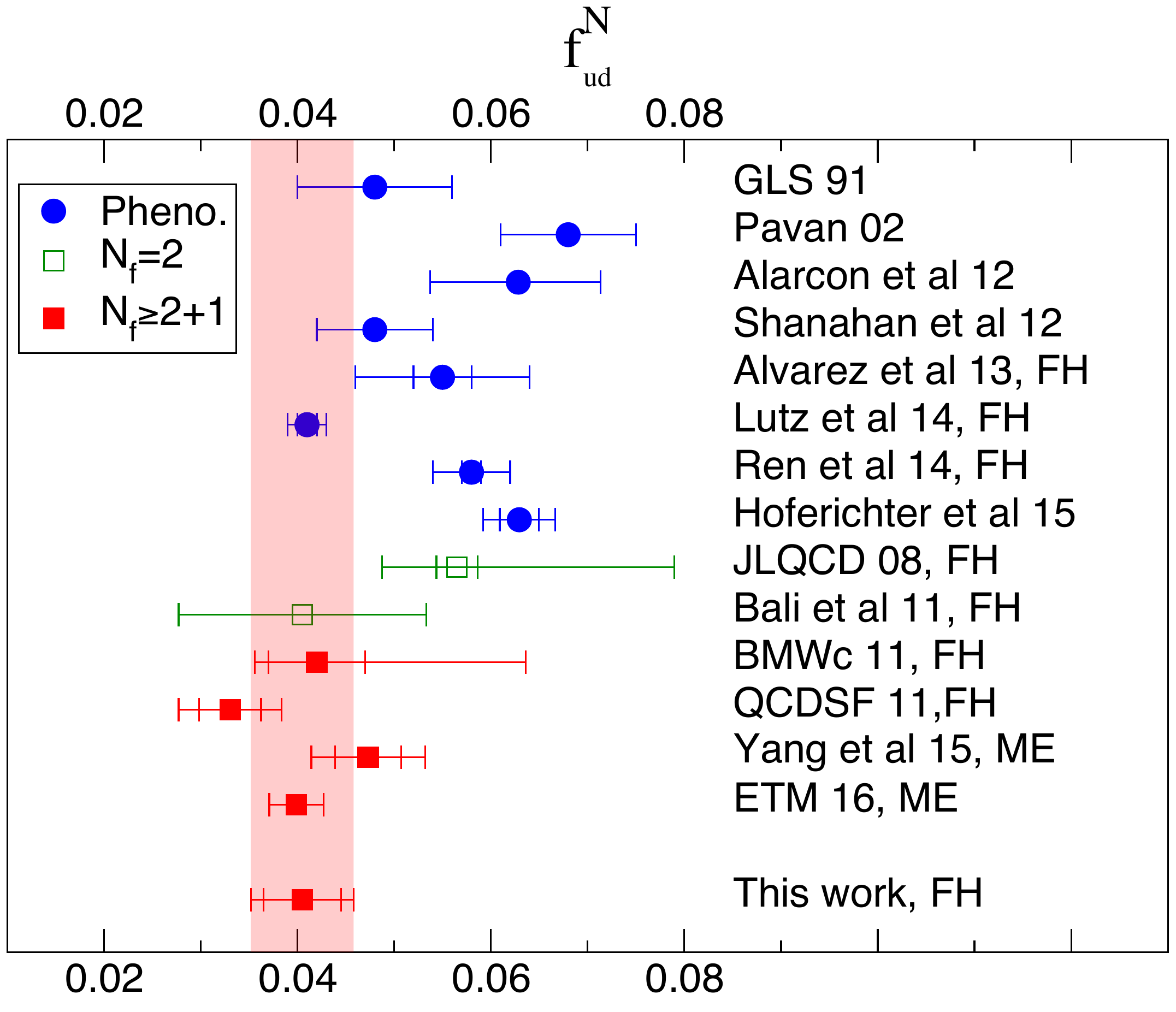}
\includegraphics[width=0.48\textwidth]{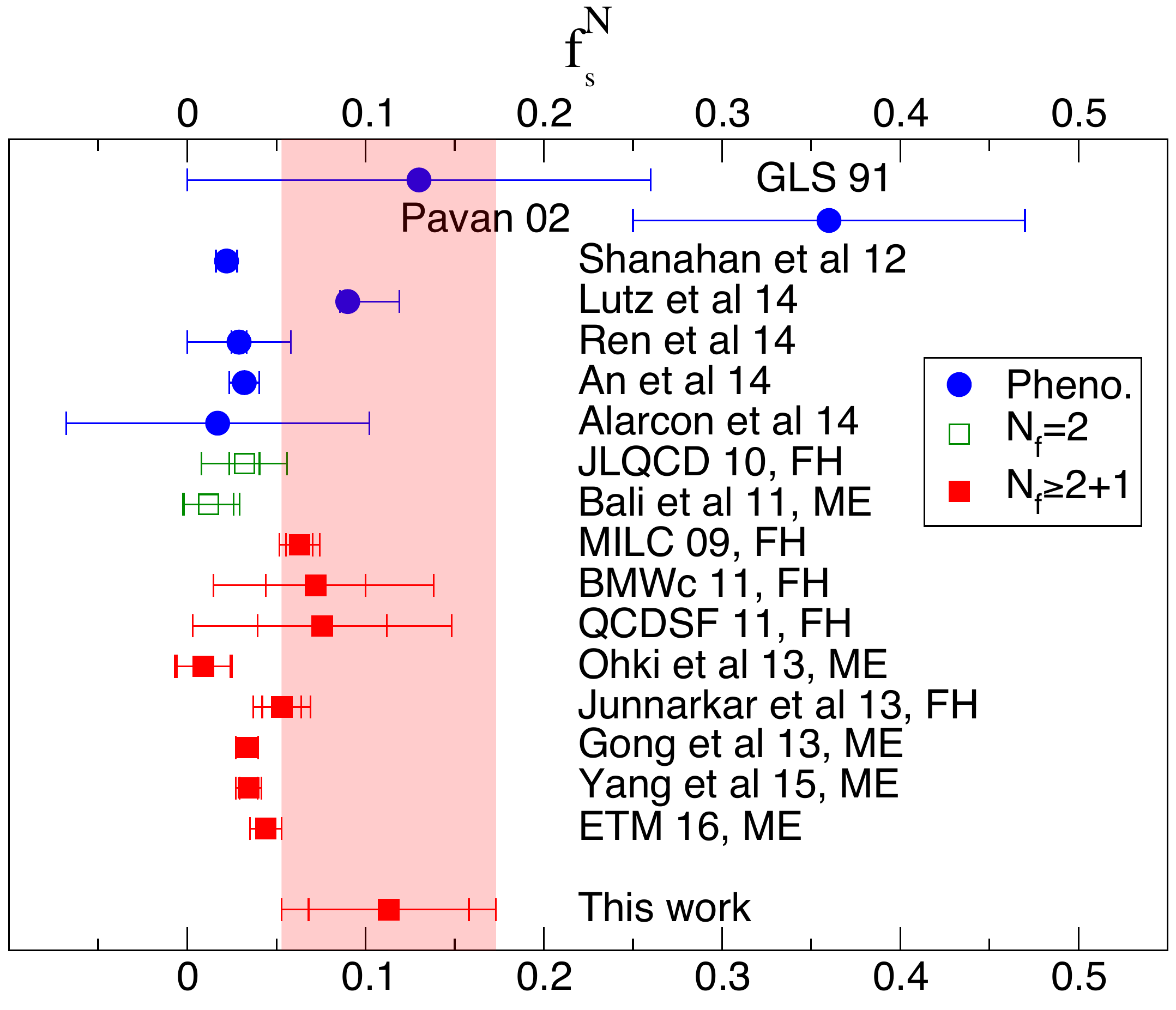}
\caption{(Color online) Comparison of our results for $f_{ud}^N$ (left panel) and for $f_{s}^N$ (right panel) with values from the literature.
Numbers are from
\cite{Gasser:1990ce} (GLS 91),
\cite{Pavan:2001wz} (Pavan 02),
\cite{Alarcon:2011zs} (Alarcon et al 12),
\cite{Shanahan:2012wh} (Shanahan et al 12),
\cite{Alvarez-Ruso:2013fza} (Alvarez et al 13),
\cite{Lutz:2014oxa} (Lutz et al 14),
\cite{Ren:2014vea} (Ren et al 14),
\cite{Hoferichter:2015dsa} (Hoferichter et al 15),
\cite{Ohki:2008ge} (JLQCD 08),
\cite{Bali:2011ks} (Bali et al 11),
\cite{Durr:2011mp} (BMWc 11),
\cite{Horsley:2011wr} (QCDSF 11),
\cite{Yang:2015uis} (Yang et al 15),
\cite{Abdel-Rehim:2016won} (ETM 16),
\cite{Gong:2013vja} ($\chi$QCD 13),
\cite{An:2014aea} (An et al 14),
\cite{Alarcon:2012nr} (Alarcon et al 14),
\cite{Toussaint:2009pz} (MILC 09),
\cite{Takeda:2010cw} (JLQCD 10),
\cite{Oksuzian:2012rzb} (Okhi et al 13),
\cite{Junnarkar:2013ac} (Junnarkar et al 13). For lattice based determinations ``FH'' denotes studies that use the Feynman-Hellmann theorem while ``ME'' denotes direct computations of the matrix element.
}
\label{fudN}
\end{figure*}

Early determinations of $\sigma_{\pi
  N}$~\cite{Koch:1982pu,Gasser:1990ce,Pavan:2001wz} were obtained
using $\pi$-$N$ scattering data. They rely on a difficult
extrapolation of the amplitude to the unphysical Cheng-Dashen point,
where small $SU(2)$ chiral perturbation theory ($\chi$PT) corrections
\cite{Gasser:1987rb,Bernard:1996nu,Gasser:1990ap,Becher:1999he,Hoferichter:2015dsa,Hoferichter:2015hva}
can be applied to obtain $\sigma_{\pi N}$. The two results
\cite{Gasser:1990ce,Pavan:2001wz} differ by nearly two standard
deviations and a factor of about 1.4, for reasons discussed in
\cite{Hoferichter:2015dsa,Hoferichter:2015hva,Alarcon:2011zs}. $\sigma_{sN}$
is then obtained from $\sigma_{\pi N}$ using results for $m_s/m_{ud}$
and $SU(3)$ $\chi$PT
\cite{Gasser:1980sb,Borasoy:1996bx,Borasoy:1998uu}. Propagating the
two determinations of $\sigma_{\pi N}$ leads to a factor of 3
difference in $\sigma_{sN}$ at the 1.5 $\sigma$ level, which gets
squared in DM-nucleon cross sections. This situation has prompted new
phenomenological and model studies (some using published lattice
results)~\cite{Hoferichter:2015dsa,Alarcon:2011zs,Shanahan:2012wh,Alvarez-Ruso:2013fza,An:2014aea,Alarcon:2012nr,Ren:2014vea,Lutz:2014oxa}
as well as a number of lattice
calculations~\cite{Alexandrou:2009qu,Toussaint:2009pz,Babich:2010at,Bali:2011ks,Horsley:2011wr,Durr:2011mp,Freeman:2012ry,Alexandrou:2013uwa,Junnarkar:2013ac,Gong:2013vja,
Yang:2015uis,Abdel-Rehim:2016won}
(see Fig.~\ref{fudN}). Recent critical reviews of $\sigma_{\pi N}$
can be found in \cite{Hoferichter:2015hva,Leutwyler:2015jga}.

Here we report on an ab initio, lattice QCD calculation, via the
Feynman-Hellmann (FH) theorem, of the nucleon scalar quark contents,
$f_{ud}^N$ and $f_s^N$, in the isospin limit. We also compute the four
quantities, $f_{u/d}^{p/n}$, to leading order in an expansion in
$\delta m=m_d-m_u$, assuming $m_{ud}\sim\delta m$, which is well
satisfied in nature. All of our results are accurate up to very small,
subleading isospin-breaking corrections. For $f_{u/d}^{p/n}$ this
represents a marked improvement over the standard approach
\cite{Ellis:2000ds}, which is only accurate up to much larger
$SU(3)$-flavor breaking corrections.

\emph{Numerical setup -} The dataset at the basis of this study
consists of $47$ ensembles with tree-level-improved Symanzik gauge
action and $N_f=2+1$ flavors of clover-improved Wilson quarks, the
latter featuring $2$ levels of HEX smearing \cite{Durr:2010aw}. The
ensembles are made up of approximately $13000$ configurations
altogether and, on average, around $40$ measurements for each correlator 
are performed on each configuration. The ensembles are obtained with $5$ lattice spacings $a$ (ranging
from $0.054$ fm to $0.116$ fm), lattice sizes up to $6$ fm and pion
masses, $M_{\pi}$, down to $120\,\text{MeV}$. This setup allows for a
consistent control of systematic uncertainties when reaching the
physical point ($\Phi$), i.e. when interpolating to physical
$m_{ud}^{(\Phi)}$ and $m_s^{(\Phi)}$ and extrapolating to
$a\rightarrow0$ and $V\rightarrow\infty$.

\emph{Methodology -} 
We use the FH theorem to compute the quark contents via the derivative of the nucleon mass with respect to the quark masses
\begin{equation}
\label{FH}
f_{ud}^N = \left.\frac{m_{ud}}{M_N}\frac{\partial M_N}{\partial m_{ud}}\right|_{\Phi}\ , \ \ \ \ \ f_{s}^N = \left.\frac{m_{s}}{M_N}\frac{\partial M_N}{\partial m_{s}}\right|_{\Phi}\ ,
\end{equation}
thus avoiding a computation of the 3-point functions required for a direct calculation of the matrix elements.

To determine the individual $u$ and $d$ contents of the proton and neutron, we start
from the simple algebraic identity (again, $\delta m=m_d-m_u$)
\begin{equation}
\label{fufdini}
\begin{split}
f_{u/d}^p&=\left(\frac{1}{2}\mp\frac{\delta m}{4m_{ud}}\right) f_{ud}^p\\
&+\left(\frac{1}{4}\mp\frac{m_{ud}}{2\delta m}\right)\frac{\delta m}{2M_p^2}\langle p|\bar{d}d-\bar{u}u|p\rangle\ .
\end{split}
\end{equation}
Note that the QCD Hamiltonian can be decomposed as
\begin{equation}
H=H_\text{iso}+H_{\delta m}\ ,\quad H_{\delta m}=\frac{\delta m}{2}\int d^3x\,(\bar{d}d-\bar{u}u)\ ,
\end{equation}
where $H_\text{iso}$ denotes the full isospin symmetric component, including the $m_{ud}$ term. To leading order in $\delta m$, the shift $\delta M_N$ to the mass of $N=p$ or $n$, due to the perturbation $H_{\delta m}$, is 
\begin{equation}
\delta M_N=\frac{\langle N|H_{\delta m}|N\rangle}{\langle N\vert N\rangle} = \frac{\delta m}{4 M_N}\langle N|\bar dd-\bar uu|N\rangle\ .
\end{equation}
Moreover, in the isospin limit $M_n=M_p$ and $\langle n|\bar dd-\bar uu|n\rangle = \langle p|\bar uu-\bar dd|p\rangle$, so that, up to higher-order isospin-breaking corrections, the $n$-$p$ mass difference is
\begin{equation}
\Delta_\text{QCD} M_N = 2\delta M_p = \frac{\delta m}{2M_p}\langle p|\bar uu-\bar dd|p\rangle\ .
\end{equation}
Using this relation, introducing the quark-mass ratio $r=m_u/m_d$ and remembering that, in the isospin limit, $f_{ud}^p=f_{ud}^N$, we obtain
\begin{equation}
\label{fudpfin}
\begin{split}
f_u^{p/n}&=\left(\frac{r}{1+r}\right) f^N_{ud}\pm\frac{1}{2}\left(\frac{r}{1-r}\right) \frac{\Delta_\text{QCD}M_N}{M_N}\ ,\\
f_d^{p/n}&=\left(\frac{1}{1+r}\right) f^N_{ud}\mp\frac{1}{2}\left(\frac{1}{1-r}\right) 
\frac{\Delta_\text{QCD}M_N}{M_N}\ ,
\end{split}
\end{equation}
where the upper sign is for $p$ and the lower one for $n$ and where
$M_N=M_n=M_p$ is the nucleon mass in the isospin limit. These
equations hold up to very small $O(\delta m^2,m_{ud}\delta m)$
corrections. Analogous expressions were obtained independently in
\cite{Crivellin:2013ipa}, using $SU(2)$ $\chi$PT.

\emph{Extracting hadron and quark masses -} Quark and hadron masses
are extracted as detailed in \cite{Durr:2010aw}, with the quark masses
determined using the ``ratio-difference method''. In addition, to
reliably eliminate excited state effects in hadron correlators we have
used a procedure similar to that suggested in
\cite{Borsanyi:2014jba}. For each of our four hadronic channels, we
fit the corresponding correlator $C(t)$ to a single-state ansatz. We
use the same minimal start time for our fit interval, $t_\text{min}$,
and the same maximum plateau length, $\Delta t$, for all ensembles: $t_\text{min}$ and
$\Delta t$ are fixed in physical and lattice units respectively. They are
determined by requiring that the distribution of fit qualities over
our 47 ensembles be compatible with a uniform distribution to better
than 30\%, as given by a Kolmogorov-Smirnov (KS) test. An identical
procedure is followed to determine the axial Ward identity (AWI)
masses.

\emph{Computing physical observables -} On each of our 47 ensembles,
we extract $M_\pi$, $M_K$, $M_N$ and $M_\Omega$ as well as the light
and strange quark masses, $m_{ud}$ and $m_s$, as explained in the last
paragraph. We define $M^2_{K^{\chi}}=M_K^2-M_\pi^2/2$ and work in a
massless scheme so the lattice spacings $a$ depend only on the
coupling $\beta$. These lattice spacings, together with all other
quantities, are determined from a global combined fit of the form
\begin{equation}
\label{Taylor}
\begin{split}
M_X^{n_X}=&\left(1+g_X^\text{a}(a)\right) \left(1+g_X^\text{FV}(M_\pi,L)\right)(M_X^{(\Phi)})^{n_X}\times\\
&\left(1+ c_X^{\text{a},ud}(a)\tilde{m}_{ud}
+c_X^{\text{a},s}(a)\tilde{m}_s + \text{h.o.t.}\right)\ ,
\end{split}
\end{equation}
for $M_X=(aM_X)/a(\beta)$, with $X=N,\,\Omega,\,\pi,\,K_\chi$,
$n_X=2$ for $X=\pi$ and 1 otherwise and where $(aM_X)$ is the hadron
mass in lattice units, as determined on a single ensemble.  The quark
mass terms, renormalized in the RGI scheme, are defined as
\begin{equation}
\tilde{m}_q=m_q^\text{RGI}-m_q^{(\Phi)}\ ,\quad m_q^\text{RGI}=\frac{(am_q)}{aZ_S\left(1+g_q^\text{a}(a)\right)}\ ,
\end{equation}
with renormalization constants $Z_S$ from \cite{Durr:2010aw}. By
h.o.t.\ we denote higher-order terms in the mass Taylor expansions, and
$g^\text{a}_X(a)$ parametrizes the continuum extrapolation of
$M_X^{(\Phi)}$, while $g^\text{FV}_X(M_\pi,L)$ parametrizes its
finite-volume corrections, according to
\cite{Colangelo:2005gd,Colangelo:2005cg}. The $c_X^{\text{a},q}(a)$,
$q=ud,s$, are equal to $c_X^{q}(1+g_X^{\text{a},q}(a))$, where the
$g_X^{\text{a},q}(a)$ parametrize the continuum extrapolation of the
slope parameters $c_X^{q}$. We define the physical point via $M_\pi$,
$M_{K^\chi}$ and $M_\Omega$. Thus, for those quantities,
$M_X^{(\Phi)}$ are fixed to $134.8\,\mev$, $484.9\,\mev$
\cite{Aoki:2013ldr} and $1672.45\,\mev$ \cite{Agashe:2014kda},
respectively.  Moreover, the corresponding discretization terms,
$g_X^\text{a}(a)$, vanish by definition, leaving only
$g_N^\text{a}(a)$.

\begin{figure*}[ht!]
\includegraphics*[width=0.48\textwidth]{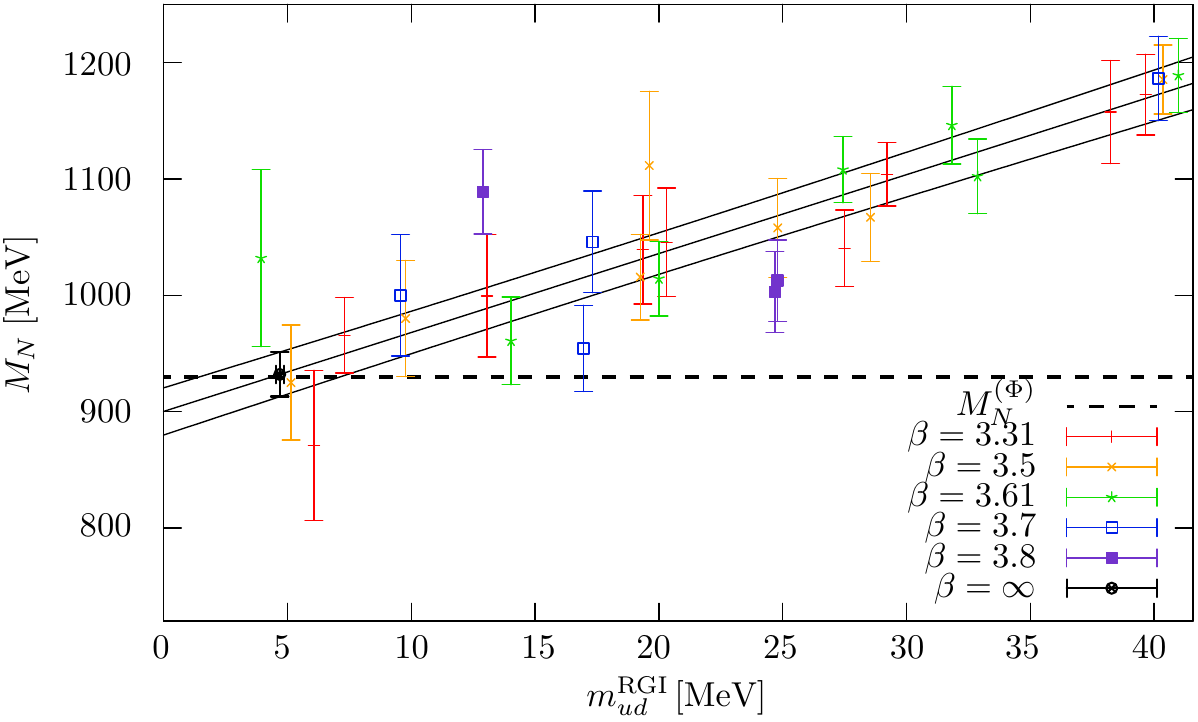}
\includegraphics*[width=0.48\textwidth]{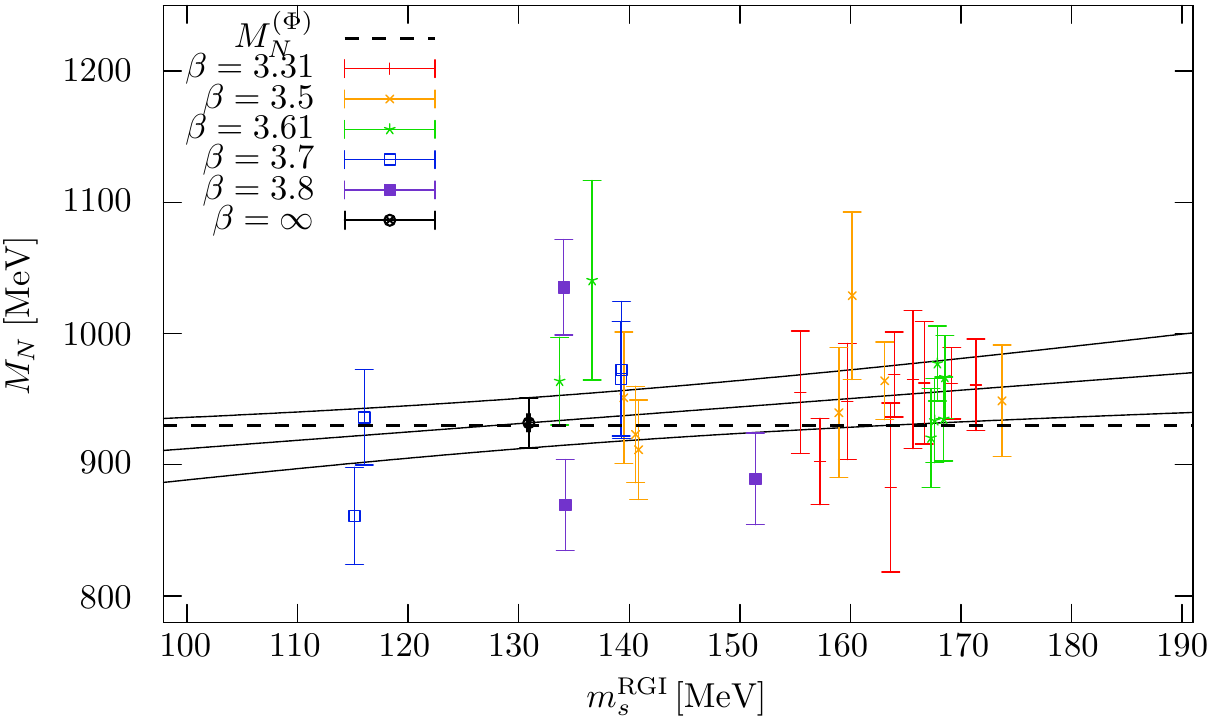}
\caption{(Color online) Typical dependence of $M_N$ on $m_{ud}^\text{RGI}$ (left
  panel) and $m_{s}^\text{RGI}$ (right panel). The black open circle
  represents our result for $M_N^{(\Phi)}$ in this particular fit,
  while the horizontal line corresponds to its experimental
  value. Dependencies of $M_N$ on variables not shown in these plots
  have been eliminated using the function obtained in the fit.}
\label{MN_vs_mud}
\end{figure*} 

\emph{Statistical and systematic uncertainties -} To estimate
systematic errors, we follow the extended frequentist method developed
in \cite{Durr:2008zz,Borsanyi:2014jba}. To account for remnant,
excited-state contributions in correlator and AWI mass fits, in
addition to the time range $(t_{min}, \Delta t)$, obtained through the
KS test, we consider a more conservative range with $t_{min}$
increased by $0.1\,\text{fm}$, while keeping $\Delta t$ fixed.
Truncation errors in the $m_{ud}$ Taylor expansion are estimated by
pruning the data with two cuts in pion mass, at $320\,\text{MeV}$ and
$480\,\text{MeV}$. In addition, we consider higher-order terms
proportional to $\tilde{m}_{ud}^2$ (or also a $\chi$PT-inspired
$[(m_{ud}^\text{RGI})^{3/2}-(m_{ud}^{(\Phi)})^{3/2}]$ for $M_N$),
$\tilde{m}_{ud}^3$, $\tilde{m}_{ud}\tilde{m}_s$ and $\tilde{m}_{ud}^2\tilde{m}_s$ in (\ref{Taylor}).
Systematic effects from terms of even higher order, which our results
are not accurate enough to resolve, are estimated by replacing the
Taylor expansions that include higher-order terms with their inverse,
in the spirit of Pad\'e approximants.  Regarding cutoff effects, our
action formally has leading corrections of $\emph{O}(\alpha_s a)$,
which are often numerically suppressed by HEX smearing, leaving a
dominant $\emph{O}(a^2)$ term \cite{Durr:2008rw}. We estimate the
uncertainty associated with the continuum extrapolation of the leading
$M_N^{(\Phi)}$ term in (\ref{Taylor}) by allowing $g_N^\text{a}(a)$ to
be proportional to either $\alpha_s a$ or $a^2$. Moreover, so as not
to over-fit the lattice results, we neglect corrections whose
coefficients are larger than 100\% except for the ones proportional to 
$\tilde{m}_{ud}^2$ (or $[(m_{ud}^\text{RGI})^{3/2}-(m_{ud}^{(\Phi)})^{3/2}]$) in $M_N$.

This procedure leads to $192$ different analyses, each one providing a
result for the observables of interest. Our final results are obtained
by weighting these $192$ values with Akaike's Information Criterion
(AIC), the AIC-weighed mean and standard deviation corresponding to
the central value and systematic error of the given observable,
respectively~\cite{Borsanyi:2014jba}. The statistical error is then
the bootstrap error of the AIC-weighted mean. The results were
crosschecked by replacing the AIC weight with either a uniform weight
or a weight proportional to the quality of each fit. In both cases we
obtained consistent values for all of our observables.

The results thus obtained account for uncertainties associated with
the continuum extrapolation of the leading $M_N^{(\Phi)}$ term in
(\ref{Taylor}), but not of the sub-leading $f_{ud}^N$ or even smaller
contribution, $f_s^N$. Indeed, the discretization terms,
$g_N^{\text{a},q}(a)$, were set to zero in the above analyses. To
account for those uncertainties, we allow the terms
$g_N^{\text{a},q}(a)$ to be proportional to $\alpha_s a$ or $a^2$. To
stabilize the corresponding fits, we fix $M_N^{(\Phi)}$ to its
experimental value. Even then, we find that within their statistical
errors, our results only support discretization corrections in
$c_N^{\text{a},ud}(a)$. Including these, and performing the same
variation of 192 analyses as in the procedure described above, we find
that the central value of $f_{ud}^N$ increases by $0.0024$ and $f_s^N$
decreases by $0.038$ compared to our standard analysis.\footnote{At
  first sight one may be surprised by the fact that adding
  discretization terms to $f_{ud}^N$ has a larger effect on
  $f_s^N$. However, these terms are small corrections to the $ud$-mass
  dependence of $M_N$ in the range of masses considered, but they are
  of similar size to the $s$-mass dependence of $M_N$ and interfere
  with it. We expect continuum extrapolation error on $f_s^N$ to be
  much smaller than the variation observed here and therefore consider
  this variation to be a conservative estimate of continuum extrapolation
  uncertainties.} We take this variation to be our estimate of the
uncertainty associated with the continuum extrapolation of the quark
contents, add it in quadrature to the systematic error obtained in our
standard analysis and propagate it throughout.

\emph{Results and discussion -} The fit qualities in this study are
acceptable, with an average $\chi^2/\text{d.o.f.} = 1.4$. Since we do
not use the nucleon for scale setting, its physical value constitutes
a valuable crosscheck of our procedure. Of course, here we only use
the results of our standard analysis, in which this mass is a free
parameter.  We obtain $M_N = 929(16)(7)\,\text{MeV}$, which is in
excellent agreement with the isospin averaged physical value
$938.9\,\text{MeV}$, as obtained by averaging $p$ and $n$ masses from
\cite{Agashe:2014kda}. Typical examples of the dependence of $M_N$ on
$m_{ud}^\text{RGI}$ and $m_s^\text{RGI}$ are shown in
Fig.~\ref{MN_vs_mud}.

The final results for the isospin-symmetric, scalar quark contents are
\begin{equation}
f_{ud}^N = 0.0405(40)(35)\ ,
\qquad
f_{s}^N = 0.113(45)(40)\ .
\label{results}
\end{equation}
Here the systematic error includes the continuum extrapolation
uncertainty discussed in the preceding section. As a further
crosscheck, we have performed an additional, full analysis where we
replace, for each lattice spacing, the renormalized quark masses in
(\ref{Taylor}) by the ratio of the lattice quark masses to their
values at the physical mass point. In this analysis, the need for
renormalization factors is obviated, because they cancel in the
ratios. However, eight additional parameters are required: at each of
our five lattice spacings, two parameters are needed to specify the
values of the $ud$ and $s$ quark masses corresponding to the physical
mass point, while only the two parameters $m_q^{(\Phi)}$, $q=ud,\,s$,
of (\ref{Taylor}) are needed for our standard analysis. Nevertheless,
the results obtained with this alternative approach are in excellent
agreement with the results from our main strategy.

It is straightforward to translate the results of Eq.~(\ref{results})
into $\sigma$ terms. We obtain $\sigma_{\pi N}=38(3)(3)\,\text{MeV}$ and
$\sigma_{sN}=105(41)(37)\,\text{MeV}$.  Another quantity of interest
in that context is the so-called strangeness content of the nucleon,
$y_N=2\langle N|\bar ss|N\rangle/\langle N|\bar uu+\bar dd|N\rangle$,
that we obtain with $m_s/m_{ud}$ determined self-consistently in our
calculation. Our result is $y_N=0.20(8)(8)$.

Now, using (\ref{fudpfin}), together with the result for $f_{ud}^N$,
the strong isospin splitting of the nucleon mass, $\Delta_\text{QCD}
M_N=2.52(17)(24)\,\text{MeV}$ from \cite{Borsanyi:2014jba}, and the
quark-mass ratio $r=0.46(2)(2)$ from \cite{Aoki:2013ldr}, we find 
\begin{equation}
\begin{array}{rclcrcl}
  f_u^p &=& 0.0139(13)(12)\ ,
&\quad&
  f_d^p &=& 0.0253(28)(24)\ ,\\[0.2cm]
  f_u^n &=& 0.0116(13)(11)\ ,
&\quad&
f_d^n &=& 0.0302(28)(25)\ .
\end{array}
\end{equation}
Another interesting quantity is $z_N\equiv\langle p|\bar
uu|p\rangle/\langle p|\bar dd|p\rangle=\langle n|\bar
dd|n\rangle/\langle n|\bar uu|n\rangle$, where the last equality holds
in the isospin limit. We find it to be $z_N=1.20(3)(3)$. This is
significantly smaller than the value of 2 that one would obtain if the
scalar densities $\bar uu$ and $\bar dd$ were replaced by the number
density operators, $\bar u\gamma^0u$ and $\bar d\gamma^0d$.

We compare our results for $f_{ud}^N$ and $f_s^N$ to phenomenological
and lattice findings in Fig.~\ref{fudN}. Our result for $f_{ud}^N$
points to a rather low value, for instance compared to the recent,
precise, phenomenological determination of
\cite{Hoferichter:2015dsa}.  Regarding $f_s^N$, our value is typically
larger than most other lattice results. Note that our error bars are
not smaller than those of all previous lattice based
calculations. However, unlike previous calculations, ours are
performed directly at that physical mass point and do not require
uncertain extrapolations to physical $m_{ud}$, nor do they make use of
$SU(3)$ $\chi$PT (e.g. in replacing $m_s$ by $M_{K^\chi}^2$ in the
definition of $f_s^N$ or in constraining the $m_s$-dependence of $M_N$
with the $m_{ud}$ and $m_s$ dependence of the baryon octet), whose
systematic errors are difficult to estimate. Given this full
model-independence, our total $13\%$ error on $f_{ud}^N$ is quite
satisfactory. Unfortunately, the overall uncertainty on $f_{s}^N$ is
still large, at $53\%$. The reason for this lies in the small
$m_s$ dependence of $M_N$, as shown in Fig.~\ref{MN_vs_mud}, which is
a major drawback of the present approach based on the FH theorem. To
try to improve on the precision, the whole analysis has also been
carried out by fixing $M_{N}^{(\Phi)}$ to its experimental
value. However, the impact on the central values and error bars is
small and therefore we do not retain this approach for our main
analysis. To our understanding, in the FH approach the uncertainty on
$f_{s}^N$ can be narrowed only by reducing the statistical error on
the data and by increasing the lever arm on $m_s$.

C.~H., L.~L. and A.~P. acknowledge the warm welcome of the Benasque Center
for Science and L.~L. of U.C. Santa Barbara's KITP (under National
Science Foundation Grant No. NSF PHY11-25915), where this work was
partly completed. Computations were performed using the JUGENE installation 
of Forschungszentrum (FZ) J\"ulich and HPC resources provided by the 
``Grand \'equipement national de calcul intensif'' (GENCI) at the 
``Institut du d\'eveloppement et des ressources en informatique scientifique'' (IDRIS) 
(GENCI-IDRIS Grant No. 52275), as well as further resources at FZ J\"ulich 
and clusters at Wuppertal and CPT. This work was supported in part by the OCEVU Labex  
(ANR-11-LABX-0060) and the A*MIDEX project (ANR-11-IDEX-0001-02) which are 
funded by the ``Investissements d'Avenir'' French government program and 
managed by the ``Agence nationale de la recherche'' (ANR) and by DFG Grant No. SFB/TRR-55.

\bibliography{refs}{}
\bibliographystyle{apsrev4-1} 

\end{document}